\documentclass[trackchanges]{aastex701}

\usepackage{amsmath}
\usepackage{comment}
\begin{document}
\title{On an empirical method to build near-equilibrium axisymmetric and triaxial galaxy models}
\author[orcid=0000-0001-5651-7779,sname='Di Cintio']{Pierfrancesco Di Cintio}
\affiliation{Istituto dei Sistemi Complessi, Consiglio Nazionale delle Ricerche, Via
Madonna del Piano 17, Sesto Fiorentino, 50019, Italy}
\affiliation{Osservatorio Astrofisico di Arcetri, Istituto Nazionale di Astrofisica,
Largo Enrico Fermi 5, Firenze, 50025, Italy}
\affiliation{Istituto nazionale di Fisica Nucleare, Via G. Sansone 1, Sesto
Fiorentino, 50019, Italy.}
\email[show]{pierfrancesco.dicintio@cnr.it}  
\author[orcid=0000-0001-5371-3432,gname=Alessandro Alberto, sname='Trani']{Alessandro Alberto Trani} 
\affiliation{Department of Astronomy, University of Concepcion, Avenida Esteban
Iturra s/n Casilla, Conception, 160-C, Chile}
\affiliation{Istituto nazionale di Fisica Nucleare, via Valerio 2, Trieste, 34127, Italy}
\email{aatrani@gmail.com}
\author[0000-0003-1119-4237,sname=Cesare,gname=Valentina]{Valentina Cesare}
\affiliation{Istituto di Radioastronomia, Istituto Nazionale di Astrofisica, Via
Gobetti 101, Bologna, 40129, Italy}
\email{valentina.cesare@inaf.it}

\begin{abstract}
We present a numerical technique to implement $N$-body realizations of axisymmetric and triaxial self-gravitating system starting from a spherical self-consistent models for which a phase-space distribution is known. The method builds up on the so-called adiabatic squeezing technique by \cite{2001ApJ...549..862H} and ideally allows to control the effective ellipticities of the particle distribution. We discuss some preliminary tests and sketch a possible improvement of our method.
\end{abstract}


\section{Introduction} 
\label{sec1}
Constructing invariant phase-space distribution functions $f$ for density potential pairs $(\rho,\Phi)$ with triaxial symmetry is still one of the currently open problems of stellar dynamics. For spherically and axially systems $f$ is recovered self consistently as a function of the conserved quantities along one particle orbits using techniques based on the inversion of the collisionless Boltzmann equation (see \citealt{1916MNRAS..76..572E,1979SvAL....5...42O,1985AJ.....90.1027M,1991MNRAS.253..414C,1993MNRAS.262..401H,2007CeMDA..97..249J}, see also \citealt{Lacroix_2018} for an extensive review). In triaxial galaxies the only quantity conserved along each  trajectory is the specific energy $\mathcal{E}=v^2/2+\Phi$. Moreover, it is well known that even for small departures from the spherical symmetry, the associated potentials $\Phi$ admit some measure of chaotic orbits. It is possible to build $f$ in closed form for a particular class of models with St\"ackel potentials, for which three independent integrals of motion are expressible in ellipsoidal coordinates (\citealt{1995NYASA.751...76H}). However, the supporting triaxial densities are often not realistic.\\
\indent $N$-body realizations of a triaxial system can be constructed using numerical schemes. For example, in the \cite{1979ApJ...232..236S} method one propagates independen orbits densely sampling the model's density profile. For each trajectory the occupation probability of a given region of phase-space is recorded, so that one its left with combined histograms of position and velocity for each specific initial condition. When generating an initial position $x,y,z$ the velocity is then sorted from the previously tabulated empirical distributions. This method was later improved to accommodate density distributions with different power-law or flat cores (\citealt{2007MNRAS.377..855T,2008MNRAS.385..647V}), multi-component systems (\citealt{2007ApJ...666..165C}), rotating bar components (\citealt{2020ApJ...889...39V}) as well as central massive objects (\citealt{2021ApJS..254...25Q}).\\
\indent \cite{2001ApJ...549..862H,2002ApJ...567..817H} proposed another empirical technique dubbed adiabatic squeezing that deforms an $N$-body realization of a spherical equilibrium, by applying to the particles an effective friction along one or two coordinate axes for an extended transient during an $N$-body integration. In practice each particle's equation of motion is augmented by a drag term of the form $-\nu_t\mathbf{v}$ where $\nu_t$ is the time friction coefficient defined up to $t=t_s$ as
\begin{equation}\label{etaKHB}
\nu_t=\nu_0\left[3\left(\frac{t}{t_s}\right)^2-2\left(\frac{t}{t_s}\right)^3\right].
\end{equation} 
After a transient $\nu$ is exponentially damped to 0 with
with the attenuation factor $\mathcal{A}(t_{\rm decay})=\exp(-t^2/2t_{\rm decay}^2)$, where $t_{\rm decay}$ is chosen to be a factor roughly 3 times larger than the unperturbed system's dynamical time $t_{\rm dyn}=\sqrt{r_c^3/GM}$. Whereas the circularized density profile at the end of the squeezing are nicely fitted by the same functional form of the original spherically symmetric density profile, the technique does not allow to dictate a priori the values of the two axial ratios $c/a$ and $b/a$.\\
\indent \cite{2025A&A...698A..28S} when studying the phase-space diffusion of ensembles of independent orbits in external potentials subjected to noise and asymptotically approaching a stationary form, found that the distributions of energy and spatial density of the tracers matched closely those of the orbits sampled from the distribution function supporting the asymptotic model. Building on these unrelated results, in this preparatory work we introduce an alternative adiabatic squeezing protocol that would allow to control the axial ratios of the deformed model.
\section{The new method}\label{method}
At variance with the original adiabatic squeezing, we do not introduce an external drag in a live $N$-body run. We drive instead a system of $N$ independent tracers sampled from a spherical equilibrium $\left[\rho(r),\Phi(r)\right]$ for which the equilibrium $f$ is known either analytically or numerically, by deforming the supporting potential while applying an additional Ornstein-Uhlenbeck process to each individual orbit \citep{Ornstein1919,1930PhRv...36..823U}. Knowing the triaxial (or axi-symmetric) generalization of $\rho(r)$, where $r$ is substituted by the ellipsoidal radius $m$ defined as
\begin{equation}
m=a\left(\frac{x^2}{a^2}+\frac{y^2}{b^2}+\frac{z^2}{c^2}\right)^{1/2};\quad a\geq b\geq c,
\end{equation}
the associated potential is obtained either analytically or solving numerically the Poisson equation. The time dependence of density and potential is contained in the intermediate and minor semi-axes $b$ and $c$. At time $t=0$, by definition, $a=b=c$ and both $\rho$ and its associated potential $\Phi$ are functions of the radial coordinate $r$.\\
\indent As a function of time, the semi-axes $s_{i}=b;c$ are adiabatically reduced from their initial values $s_{i,0}$ to the desired ones $s_{i,f}$ computing
\begin{equation}\label{eq:erf}
s_i(t)=s_{i,0}+(s_{i,f}-s_{i,0}){\rm Erf}\left[(t/\tau)^2\right],
\end{equation}
where ${\rm Erf(x)}$ is the standard error function and $\tau$ is the chosen deformation time scale. The $N$ particles are propagated with 
\begin{equation}\label{eq:OM}
\ddot{\mathbf{r}}_i =-\nabla\Phi(\mathbf{r},t)-\nu\,\mathbf{v}_i+\delta \mathbf{F}_i,
\end{equation}
where $\nu$ is a friction coefficient and $\delta F$ is a stochastic force sampled from a homogeneous 3D Gaussian distribution of unitary dispersion and renormalized with the factor
\begin{equation}\label{eq:normfac}
\Upsilon =\sqrt{\mathcal{E}\,\nu/t_c},
\end{equation}
where $t_c$ is the autocorrelation time. In this work we assume 
\begin{equation}
\nu=\frac{4\pi mG^2\rho(r)\log\Lambda}{\sigma^3(r)}.
\end{equation}
During the deformation  Eq.~(\ref{eq:OM}) is solved using the \cite{2004PhRvE..69d1107M} modified second order leapfrog scheme with constant $\Delta$ discussed in \cite{2025A&A...698A..28S}. We assume that $\delta f_i$ is a coloured noise (i.e. a stochastic process with a finite, nonzero autocorrelation time $t_c$ of the order of a crossing time of the unperturbed model, e.g., see \citealt{2003astro.ph.12434T,2004ApJ...602..678S,2025A&A...703A...6T}). The stochastic force $\delta F_i$ is updated at the $n+1$ step simulation step as
\begin{equation}
\delta F_i^{n+1}=\exp(-\Delta t/t_c)\delta F_i^n+u\hat{F}_i^{n+1},
\end{equation}
where $u=\sqrt{1-\exp(-2\Delta t/t_c)}$ and $\hat{F}_i^{n+1}$ is a new independent sampling of the noise distribution. Fixing $t_c=\Delta t$ yield a delta correlated noise.

Once both $b$ and $c$ have reached the prescribed values, the system is propagated further under the action of the external triaxial potential for a transient of the order $3\tau$ before the self consistent potential is activated.
\begin{figure}
    \centering
    \includegraphics[width=0.55\textwidth]{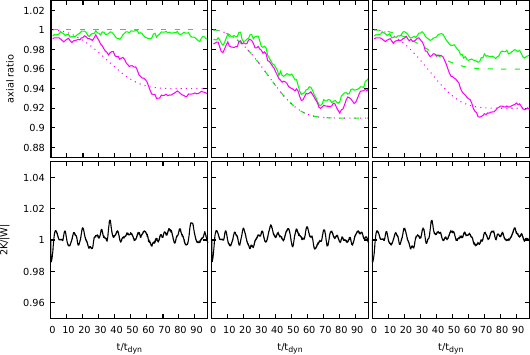}
     \includegraphics[width=0.38\textwidth]{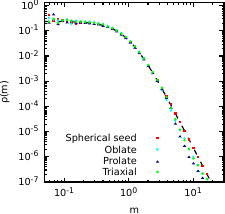} 
   \caption{Left panel, top: evolution of the axial ratios $b/a$ (green) lines and $c/a$ (magenta). The dashed lines refer to the time-evolved ratios according to Eq. (\ref{eq:erf}) while the solid lines are to the values computed for the time-dependent particle distributions. From left to right the final system has an oblate ($\epsilon=1$; $\eta=0.06$), prolate ($\epsilon=\eta=0.09$) and triaxial ($\epsilon=0.04$; $\eta=0.08$) shape. Left panel, bottom: evolution of the virial ratio. Right panel: profile for an oblate (cyan circles), a prolate (blue triangles) and a triaxial system (green diamonds) generated by adiabatically deforming a spherical initial condition (red squares) extracted by a isotropic  Plummer model. The dashed line marks the analytical density.}
    \label{fig:squeeze}
\end{figure}
\section{Results and discussion}\label{sec2}
In this work, as a test we adopt the \cite{2005A&A...437..419C} models obtained via the homeoidal expansion deforming a spherical seed density defined by 
expanding at first order in the flattening parameters $\epsilon=1-b/a$ and $\eta=1-c/a$ as
\begin{equation}\label{eq:rho1}
{\rho}(m)={\rho}({r})+\frac{\epsilon{y}^2+\eta{z}^2}{\tilde{r}}{\rho}^\prime({r})+O(\epsilon^2+\eta^2); \qquad {\rho}^\prime({r})=\left.\frac{{\rm d}{\rho}(m)}{{\rm d}m}\right|_{\epsilon=\eta=0}
\end{equation}
and
\begin{equation}\label{eq:phi1}
{\Phi}({r})={\phi}_0({r})+(\epsilon+\eta)\left[{\phi}_1({r})-{\phi}_0({r})\right]+(\epsilon{y}^2+\eta{z}^2){\phi}_2({r})+O(\epsilon^2+\eta^2).
\end{equation}
In the equations above the three auxiliary potentials are 
\begin{align}\label{potentials}
{\phi}_0({r})&=-\frac{1}{{r}}\int_{0}^{{r}}{\rho}(m)m^2{\rm d}m-\int_{{r}}^{\infty}{\rho}(m)m{\rm d}m,\nonumber\\
{\phi}_1({r})&=-\frac{1}{3{r}^3}\int_{0}^{{r}}{\rho}(m)m^4{\rm d}m-\frac{1}{3}\int_{{r}}^{\infty}{\rho}(m)m{\rm d}m,\\
{\phi}_2({r})&=\frac{1}{{r}^5}\int_{0}^{{r}}{\rho}(m)m^4{\rm d}m. \nonumber
\end{align}
\indent Using the simple \cite{1911MNRAS..71..460P} model allows to derive the expressions of $\phi_i$ as well as their gradients in terms of simple functions.  In the top (left) panels of Figure \ref{fig:squeeze} we show  the evolution of the two axial ratios $b/a$ and $c/a$ according to Eq. (\ref{eq:erf}) as well as their counterparts for a $N=10^5$ particle distributions, evaluated within the standard diagonalization procedure of the inertia tensor (e.g. see \citealt{2006MNRAS.370..681N,2013MNRAS.431.3177D}) within the time dependent Lagrangian radius enclosing the $70\%$ of the total mass $M$, for an oblate a prolate and a moderately triaxial systems implemented with the CB05 density given by Eq. (\ref{eq:rho1}). We observe that, in general, the value of the axial ratios of the particle system match approximately well their counterparts for the smooth potential model. In all cases we used the deformation time scale $\tau=50t_{\rm dyn}$ and we propagate the $N$ independent equations of motion (\ref{eq:OM}) up to $t=100t_{\rm dyn}$ with a fixed $\Delta t=0.01$. With such a choice, the oscillations of the virial ratio (plotted in the bottom panels of Fig. \ref{fig:squeeze}) $2K/|W|$, where $K$ is the total kinetic energy and $W=\sum_N\langle\mathbf{r,\nabla\Phi}\rangle$ the virial function, remain of the order of the $3\%$. In the right hand pannel we show the angle averaged density profiles of the three systems. We observe that, as expected, the functional form of $\rho(m)$ matches rather well the spherical seed density.\\
\indent We have then integrated the models obtained by adiabatic deformation for different combinations of $N$ and axial ratios  $\eta,\epsilon$ with the {\sc fvfps} code (see \citealt{2006MNRAS.370..681N}) We observe that the flattening is preserved reasonably well over roughly $100 t_{\rm dyn}$. However, for models with larger departure from the spherical symmetry, some non-negligible sphericization was observed. We interpret this as an incomplete relaxation of the particle distribution the the phase-space distribution of the target non-spherical model.\\  
\indent We have introduced an alternative version of the adiabatic squeezing technique to implement $N$-body realizations of non-spherical systems starting from a spherical seed model and applying a Ornstein-Uhlenbeck process to enhance the phase-space transport in the time dependent potential. The main improvements with respect to the original technique formulated by \cite{2002ApJ...567..817H} consists in the possibility of controlling the desired axial ratios (at least for small flattening). In this preliminary work we restricted ourselves to a simple density potental pair potential generalizing the Plummmer model using the homeoidal expansion of \cite{2005A&A...437..419C}. Of course, the method can be applied to any density potential pair provided that $\rho(m)$ can be obtained by adiabatic deformation of a spherical model, regardless of the specific values of the asymptotic axial ratios $c/a$ and $b/a$.\\
\indent We performed test $N-$body simulation using as initial conditions some of the systems obtained deforming spherical equilibria, finding that they preserve rather acceptably their structure up to $100$ dynamical times. Later, in some cases some process drives the model back to a nearly spherical symmetry for low values of $\eta, \epsilon$. No particular dependence on the particle number was observed during the deformation transient. This established, it remains to be determined how does starting from the same spherical seed density, supported by different velocity distributions, responds to the same deformation of the external smooth potential. 
\bibliography{sn-bibliography}{}
\bibliographystyle{aasjournalv7}
\end{document}